\documentclass[12pt,a4paper]{article}
\usepackage{amsmath}
\usepackage{subcaption}
\usepackage{amssymb}
\usepackage[numbers]{natbib}
\usepackage{authblk}
\usepackage{epsfig}
\usepackage{graphicx}
\usepackage{float}
\usepackage{IEEEtrantools}
\usepackage{indentfirst}

\begin{document}
\title{Role of virial coefficients in chemical reaction}
\author{P Vipin, R Sankaranarayanan}
\affil{Department of Physics, National Institute of Technology, Tiruchirappalli-620015, Tamilnadu, India.}
\date{}
\maketitle{}
\begin{abstract}
	van't Hoff equation relates equilibrium constant $K$  of a chemical reaction to temperature $T$. Though the van't Hoff plot ($\ln K$ vs $1/T$) is linear, it is nonlinear for certain chemical reactions. In this work we attribute such observations to virial coefficients.
\end{abstract}
\subsection*{Introduction}
Chemical kinetics is an important subsection of physical chemistry. It deals with the rate of chemical reaction and its dependence on concentration of reactants/products and on temperature. The most widely used equation that describes the temperature dependence of reaction rate constant is the Arrhenius equation, which is proposed by Svante Arrhenius in the year 1889. In fact, Arrhenius equation is one of the three solutions of a first order differential equation which was originally proposed by Jacobus Henricus van't Hoff in the year 1884. The importance of Arrhenius' work in this regard is that he adopted and applied his solutions to a wide variety of chemical reactions \citep{laidler1985}. 
\par One can derive the van't Hoff equation from pure thermodynamical considerations on a mixture of ideal gases. The resulting equation predicts that the logarithm of equilibrium constant $K$ (ratio of rate constants associated to forward and backward reactions) is inversely proportional to temperature $T$. However this is not the case with some chemical reactions \citep{bennett1972,field1971,sluyters2017}.  
With these results in mind, here we revisit the van't Hoff equation by relaxing the ideal gas approximation. Within the framework of van't Hoff, where the heat of the reaction is constant, it is shown that the temperature dependence of $\ln K$ appears through virial coefficients as well. We demonstrate our result with an ideal interatomic potential. 
\subsection*{Mixture of gases}
Virial equation expresses equation of state of a gas as a density expansion and is given by \citep{reichl1999}
\begin{equation}
\frac{P}{k_BT}=\frac{N}{V}+\left(\frac{N}{V}\right)^{2}B_2(T)+\left(\frac{N}{V}\right)^{3}B_3(T)+\cdots
\end{equation}
where $P$, $V$ and $T$ are the pressure, volume and temperature of the gas respectively. If $N$ is the number of molecules, for one mole of gas ($N=N_A$, Avogadro number) we have the gas constant $R = N_Ak_B$, where $k_B$ is the Boltzmann constant. Here $B_2(T)$ and $B_3(T)$ are called the second and third virial coefficients respectively. These coefficients can be calculated using intermolecular potentials through cluster expansions. 
\par 
 van't Hoff equation is a first order differential equation which describes the temperature dependence of equilibrium constant  for a mixture of gases obeying ideal gas law. In what follows, we derive van't Hoff equation using the virial equation of state. Considering one mole of the gas, and retaining upto $B_3(T)$ we have
\begin{equation}
\left(\frac{\partial{P}}{\partial{T}}\right)_V=\frac{R}{V}+\frac{N_AR}{V^{2}}f_{2}(T)+ \frac{N_{A}^{2}R}{V^3}f_3(T)
\end{equation}
where $f_{l}(T) = B_l(T)+TB'_l(T)$ with $l = 2,3$. Recollecting the first law
\begin{equation}
TdS=C_VdT+T\left(\frac {\partial{P}}{\partial{T}}\right)_VdV
\end{equation}
and upon integration, the entropy is 
\begin{equation}
S=(C_V \ln T+R \ln V)-\frac{N_AR}{V}f_{2}(T)-\frac{N_{A}^{2}R}{2V^2}f_3(T)+\alpha
\end{equation}
where $C_V$ is the specific heat at constant volume and $\alpha$ is the integration constant. With this, the Helmholtz free energy $F=U-TS$ is obtained as
\begin{equation}
F=C_VT+{\cal U}-T\left(C_V\ln T+R\ln V-\frac{N_AR}{V}f_{2}(T)-\frac{N_{A}^{2}R}{2V^2}f_3(T)+\alpha\right)
\end{equation}
where ${\cal U}$ is the energy of the gas at absolute zero.
\par If a gas contains $[A_1]$ moles per unit volume, the volume of one mole would be $1/[A_1]$. Then the free energy of $m_1$ moles of $A_1$ in volume $V$ is simply
\begin{eqnarray}
 \lefteqn{F}\, &=&V[A_1]\Big\{C_{V1}T+{\cal U}_1-T\Big(C_{V1}\ln T-R\ln [A_1]\nonumber\\&&-N_AR[A_1]f_{21}(T)-\frac{N_{A}^{2}R[A_1]^2}{2}f_{31}(T)+\alpha_1\Big)\Big\}
\end{eqnarray} 
where $C_{V1}$, ${\cal U}_1$ and $\alpha_1$ are quantities associated with the gas $A_1$. Here $f_{21}(T)$ and $f_{31}(T)$ are the terms associated with second and third virial coefficients of the gas $A_1$ respectively. We thus have similar expressions for free energies of all other gases in a gaseous mixture of volume $V,$ undergoing a reaction given by
\begin{equation}
\sum_{i=1}^{r} m_iA_i \leftrightharpoons \sum_{j=1}^{p} n_jB_j
\end{equation}
where $m_i$ and $n_j$ are integers. Assuming that total free energy of the gaseous mixture $\tilde{F}$ is still additive: 
\begin{eqnarray}
\lefteqn{ \tilde{F}}\,&=& V\sum_{i=1}^{r}[A_i]\Big\{C_{Vi}T+{\cal U}_i-T\Big(C_{Vi}\ln T-R\ln [A_i] \nonumber\\&&-N_AR[A_i]f_{2i}(T)-\frac{N_{A}^{2}R[A_i]^2}{2}f_{3i}(T)+\alpha_i\Big)\Big\}\nonumber\\&&+V\sum_{j=1}^{p}[B_j]\Big\{C_{Vj}T+{\cal U}_j-T\Big(C_{Vj}\ln T-R\ln [B_j]\nonumber\\&&-N_{A}R[B_j]f_{2j}(T)-\frac{N_{A}^{2}R[B_j]^2}{2}f_{3j}(T)+\alpha_j\Big)\Big\}
\end{eqnarray}
where $[A_i] $'s and $[B_j]$'s are the concentrations of the reactants and products respectively. Here we follow the convention that quantities with subscripts $i$ and $j$ are associated to the gases $A_i$ and $B_j$ respectively.
\subsection*{The van't Hoff equation}
Let the reaction proceeds from left to right, and infinitesimal amount of gases $A_i$, say $d[A_i]$ react to form infinitesimal amount of gas $B_j$, i.e. $d[B_j]$. To make the analysis simpler, we assume that the variations $d[A_i]$ and $d[B_j]$ are proportional to $m_i$ and $n_j$ respectively. If we take the proportionality constant to be $\epsilon > 0$, then $d[A_i] = -\epsilon m_i$ and $d[B_j] = \epsilon n_j$; negative sign signifies that the quantity $[A_i]$ decreases to form $[B_j]$. With this consideration, for the total free energy of the reaction to be minimum, we must have $d\tilde{F}=0$. That is at equilibrium
\begin{equation}
d \tilde{F}=\epsilon \left\{ -\sum_{i=1}^{r}\frac {\partial{\tilde{F}}}{\partial{[A_i]}}m_i+\sum_{j=1}^{p}\frac {\partial{\tilde{F}}}{\partial{[B_j]}}n_j \right\}=0.
\end{equation}
Since $\epsilon$ and $V$ are constants, the above condition translates to
\begin{eqnarray}
\lefteqn{RT\left(\sum_{i=1}^{r}m_i\ln [A_i]-\sum_{j=1}^{p}n_j\ln [B_j]\right)}\nonumber\\
&=&\sum_{i=1}^{r}\Big\{-m_iC_{Vi}T-m_i{\cal U}_{i}+m_iC_{Vi}T\ln T-m_iRT+m_i \alpha_iT \nonumber\\
&&-2m_i[A_i]N_{A}RTf_{2i}(T)-\frac{3}{2}m_i[A_i]N_{A}^{2}RTf_{3i}(T)\Big\} 
\nonumber\\&&+\sum_{j=1}^{p}
\Big\{n_jC_{Vj}T+n_j{\cal U}_j-n_jC_{Vj}T\ln T+n_jRT-n_j\alpha_jT\nonumber\\&&+2n_j
[B_j]N_{A}RTf_{2j}(T)+\frac{3}{2}n_j[B_j]N_{A}^{2}RTf_{3j}(T)\Big\}.
\end{eqnarray}
Dividing by $RT$, we get 
\begin{eqnarray}
 \ln K&=&-\frac{1}{R}\sum_{i=1}^{r}m_iC_{Vi}-\frac{1}{RT}\sum_{i=1}^{r}m_i{\cal U}_{i}+\frac{\ln T}{R}\sum_{i=1}^{r}m_iC_{Vi}+\frac{1}{R}\sum_{i=1}^{r}m_i \alpha_i\nonumber\\&&-\sum_{i=1}^{r}m_i -2N_{A}\sum_{i=1}^{r}m_i[A_i]f_{2i}(T)-\frac{3}{2}N_{A}^{2}\sum_{i=1}^{r}m_i[A_i]f_{3i}(T)\nonumber\\ &&{}+\frac{1}{R}\sum_{j=1}^{p}n_jC_{Vj}+\frac{1}{RT}\sum_{j=1}^{p}n_j{\cal U}_j-\frac{\ln T}{R}\sum_{j=1}^{p}n_jC_{Vj}-\frac{1}{R}\sum_{j=1}^{p}n_j \alpha_j\nonumber\\&&+\sum_{j=1}^{p}n_j
+2N_{A}\sum_{j=1}^{p}n_j[B_j]f_{2j}(T)+\frac{3}{2}N_{A}^{2}\sum_{j=1}^{p}n_j[B_j]f_{3j}(T)
\end{eqnarray}
where we have defined equilibrium constant associated to the reaction as
\begin{equation}
K = \frac{\prod_{i=1}^{r}[A_i]^{m_i}}{\prod_{j=1}^{p}[B_j]^{n_j}}.
\end{equation}
 \par Finally, differentiating with respect to temperature, we get
\begin{eqnarray}
\frac{d}{dT}\left(\ln K\right) &=&\frac{1}{RT^2}\left(\sum_{i=1}^{r}m_i{\cal U}_i- \sum_{j=1}^{p} n_j{\cal U}_j \right)+\frac{1}{RT^2}\left(\sum_{i=1}^{r}m_iC_{Vi}T - \sum_{j=1}^{p}n_jC_{Vj}T\right)\nonumber\\&&+2N_A\left\{\sum_{j=1}^{p}n_j[B_j]f'_{2j}(T)-\sum_{i=1}^{r}m_i[A_i]f'_{2i}(T)\right\}\nonumber\\&&+\frac{3}{2}N_{A}^{2}\left\{\sum_{j=1}^{p}n_j[B_j]f'_{3j}(T)-\sum_{i=1}^{r}m_i[A_i]f'_{3i}(T)\right\}.
\end{eqnarray}
Defining heat of the reaction (van't Hoff enthalpy) as
\begin{equation}
H=\sum_{i}^{r}m_i(C_{Vi}T+{\cal U}_i)-\sum_{j}^{p}n_j(C_{Vj}T+{\cal U}_j)
\end{equation}
in terms of which the reactions are classified into exothermic ($H>0$) and endothermic ($H<0$).
With this, we get the modified van't Hoff equation as
\begin{eqnarray}
 \frac{d}{dT}\left(\ln K\right) =\frac{H}{RT^2} +2N_A\left\{\sum_{j=1}^{p}n_j[B_j]f'_{2j}(T)-\sum_{i=1}^{r}m_i[A_i]f'_{2i}(T)\right\}\nonumber\\+\frac{3}{2}N_{A}^{2}\left\{\sum_{j=1}^{p}n_j[B_j]f'_{3j}(T)-\sum_{i=1}^{r}m_i[A_i]f'_{3i}(T)\right\}.
\end{eqnarray}
The solution to the above equation can be written straightforwardly as
\begin{eqnarray}
\ln K = -\frac{H}{RT}+2N_A\left\{\sum_{j=1}^{p}n_j[B_j]f_{2j}(T)-\sum_{i=1}^{r}m_i[A_i]f_{2i}(T)\right\}\nonumber\\  +\frac{3}{2}N_{A}^{2}\left\{\sum_{j=1}^{p}n_j[B_j]f_{3j}(T)-\sum_{i=1}^{r}m_i[A_i]f_{3i}(T)\right\}.
\end{eqnarray}
\par The above equation shows temperature dependent additive terms, arising out of the second and third virial coefficients, which are associated to all the constituent gases in the reacting mixture. To further simplify the terms for better analysis, we shall consider $f_{2i(j)}(T) = f_2(T)$ and $f_{3i(j)}(T) = f_3(T)$. With this the above equation becomes
\begin{equation}
\ln K = -\frac{H}{RT}+dM\left(2N_{A}f_{2}(T)+\frac{3}{2}N_{A}^{2}f_3(T)\right)
\end{equation}
 where $dM =\sum_{j=1}^{p}n_j[B_j]-\sum_{i=1}^{r}m_i[A_i]$. Thus we get the van't Hoff approximation for the equilibrium constant $K$ if $dM=0$, which is equivalent to the ideal gas approximation i.e., $B_2(T) = B_3(T) = 0$. With this form of van't Hoff equation, we now turn our attention to a simple interatomic potential for which expression for the second virial coefficient is known.  

\subsection*{Square well potential}

Let us consider an interatomic potential in the form of a square well given by
 \begin{figure*}[t]
	\begin{center}
		\includegraphics*[width=0.60\linewidth]{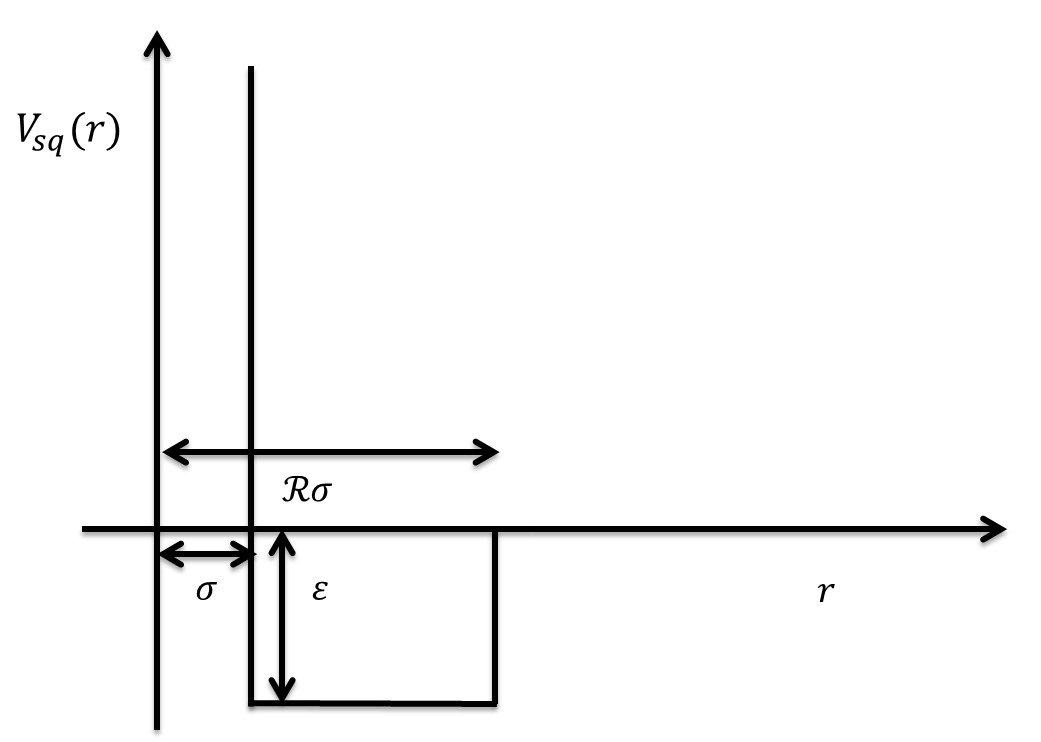}
		\caption{\small The square well potential used in the text.  The parameters chosen are ${\cal{R}}= 1.62$, $\sigma=3.27${\AA}  and $\varepsilon/k_{B}=88.3$ K which correspond to the numerical fit with the experimentally determined $B_2(T)$ for nitrogen gas.}
	\end{center}	
\end{figure*}

\begin{equation}
V_{sq}(r)=\left\{\,
				\begin{IEEEeqnarraybox}[][c]{l?s}
					\IEEEstrut
						\infty & if $0<r<\sigma$, \\
						-\varepsilon & if $\sigma<r<\cal{R}\sigma$, \\
				 0  & if ${\cal{R}}\sigma< r $.
					\IEEEstrut
				\end{IEEEeqnarraybox}
		\right. 
\end{equation}
This potential is characterized by three parameters namely $\sigma, \cal{R}$ and $\varepsilon$. While $\sigma$ signifies the distance over which the molecules exert repulsion, scaling factor $\cal{R}$ and $\varepsilon$ signify the width and depth of the well respectively - both responsible for molecular attraction. These parameters, as shown in Fig. $1$ can be obtained for a particular gas by appropriate numerical fitting with experimental values of $B_2(T)$.

\par The second virial coefficient for the square well potential having width $({\cal{R}}-1)\sigma$ is given by
\begin{equation}
B_2(T)=A-B\,\exp\,\left(\frac{\varepsilon}{k_BT}\right) 
\end{equation}
where $A = (2/3)\pi \sigma^3 {\cal{R}}^3$ and $B = (2/3)\pi \sigma^3 ({\cal{R}}^3-1)$. With this we can write
\begin{equation}
f_2(T)=A-B\,\exp\,\left(\frac{\varepsilon}{k_BT}\right)+\frac{B{\varepsilon}}{k_BT}\exp\,\left(\frac{\varepsilon}{k_BT}\right). 
\end{equation}
\begin{figure}
		\centering
		\begin{subfigure}[t]{0.48\linewidth}
		\includegraphics[width=\linewidth]{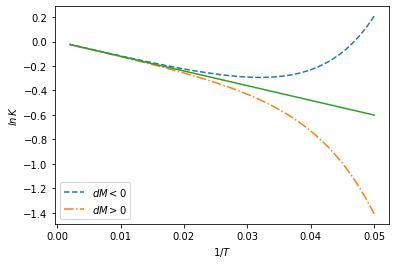}
		\caption{}
		\end{subfigure}
		\begin{subfigure}[t]{0.48\linewidth}
		\includegraphics[width=\linewidth]{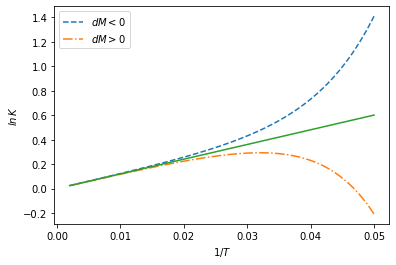}
		\caption{}
		\end{subfigure}
		\caption{\small(Colour online) Temperature dependence of equilibrium constant for (a) exothermic and (b) endothermic reactions. Solid lines correspond to the ideal gas approximation ($dM=0$).} 
\end{figure}Ignoring $B_3(T)$ and so $f_3(T)$ for the square well potential and since $dM$ in eq.$(17)$ is a constant for a given reactive mixture, the term $f_2(T)$ will be responsible for any deviation from the ideal gas.

Fig. 2 shows the van't Hoff plots for a mixture of real gases modelled by a square well potential for exothermic and endothermic reactions. In both the cases we observe that the ideal gas approximation $(dM=0)$ is valid for $T \gtrsim 65\,K$.
However, the deviations in equilibrium constant from ideal gas approximation increase as temperature decreases. The observed deviations arise from the temperature dependent virial coefficient given by eq.$(19)$.
It is also understandable that the van't Hoff plots are \textit{convex} for $dM >0$, indicating that the equilibrium constant is lower than that of ideal gas approximation at lower temperatures. Similarly, for $dM<0$ the plots are \textit{concave}, indicating that the constant $K$ is larger than the van't Hoff's prediction at lower temperatures. These kinds of behaviours are dubbed as \textit{super-Arrhenius} and \textit{sub-Arrhenius} respectively \citep{aquilanti2010temperature, silva2016}. 
 \begin{figure}
 	\centering
 		\begin{subfigure}{0.48\linewidth}
 		\includegraphics[width=\linewidth]{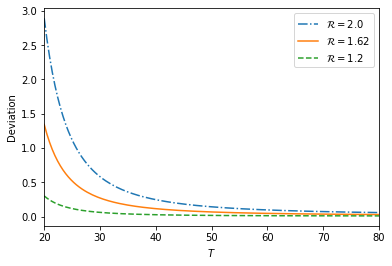} 		\caption{}
 		\end{subfigure}
 		\begin{subfigure}{0.48\linewidth}
 		\includegraphics[width=\linewidth]{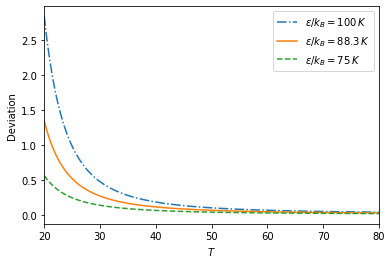}
 		\caption{}
 		\end{subfigure}
 		\caption{\small (Colour online) (a)  Deviation in van't Hoff plot with temperature for different well widths with $\varepsilon/k_B= 88.3\, K$. (b) Deviation with temperature for different well depths with ${\cal{R}} = 1.62$.}
 \end{figure}
We also observe from Fig. 3 that the deviations from ideal gas approximation at low temperature increase with the scaling factor $\cal{R}$ and depth $\varepsilon$ of the well, as expected. 

 \subsection*{Conclusion}
The traditional thermodynamic approach to derive an equation for equilibrium constant within the framework of ideal gas approximation leads to van't Hoff equation. In the present work, we have followed the same route  by relaxing the assumption that the reacting gaseous species are ideal. We find that the introduction of virial coefficients into the equation of state, give rise to general temperature dependent additive terms in the van't Hoff equation. Nonlinear behaviour in the van't Hoff plot is thus attributed to the virial coeffcients of the real gaseous mixture. The deviation of equilibrium constant from the ideal gas approximation at low temperature is demonstrated with square well potential.

\end{document}